\documentclass{epl}

\title{Memory effects in the relaxation of Ising models}
\author{J. Javier Brey \and A. Prados}
\institute{
   F\'{\i}sica Te\'{o}rica, Universidad de Sevilla, Apartado de Correos
 1065, E-41080, Sevilla, Spain
} \pacs{05.50.+q}{Lattice theory and statistics; Ising systems}
\pacs{05.70.Ln}{Nonequilibrium thermodynamics; irreversible
processes} \pacs{45.70.Cc}{Static sandpiles; granular compaction}

\begin{document}

\maketitle

\begin{abstract}
It is analytically shown that the one-dimensional Ising model with Glauber
dynamics exhibits short time memory effects when submitted to an abrupt change
in the temperature. These effects are qualitatively similar to those
experimentally observed in the compaction of vibrated granular materials.
Moreover, a critical time separating regimes of ``normal'' and ``anomalous''
responses to the perturbation is found.
\end{abstract}


The dynamics of granular systems exhibits a rich phenomenology that has been the
subject of intensive study in the last years \cite{JNyB96}. In particular, when
submitted to vertical vibration starting from a loose packing configuration,
granular media compact behaving in a way that is reminiscent of conventional
structural glasses \cite{KFLJyN95,NKBJyN98}. The glassy nature of granular
compaction clearly shows up when the response of the system to a sudden change
in the vibration intensity is measured \cite{Ni99,JTMyJ00,ByP01}. The results
show the presence of relevant short-term memory effects that are in contrast
with the long time behaviour observed at constant intensity. A decrease
(increase) in the vibration acceleration leads to an increase (decrease) of the
compaction rate at short times. For large times, the previous history is
forgotten and the relaxation rate becomes the same as for a constant intensity
process.

Here we study the existence of a similar ``anomalous'' response in the context
of the one-dimensional Ising model with Glauber dynamics \cite{Gl63}. This is a
model with short range interactions that exhibits many of the qualitative
features of conventional structural glasses, like slow non-exponential
relaxation, a kinetic phenomenon resembling a laboratory glass transition,
aging, and hysteresis effects
\cite{An70,Sk83,ByS85,Sp89,ByP93,ByP96a,KyS90,ByP94,Br90,PByS97,LyZ00,GyL00}. On
the other hand, it is simple enough to allow exact analytical calculations in
situations where more realistic models are untractable. In particular, we will
be able to calculate the short-time response of the  system to a sudden small
change in the temperature, which plays in the model a role analogous to the
vibration intensity in granular compaction. As long as the change is not made at
the early stages of the relaxation process, the system shows an ``anomalous''
response. On the other  hand, if the perturbation is introduced very soon, the
response is ``normal'', in the sense that the modification of the relaxation
rate has the same sign as the temperature change. Let us mention that these two
qualitatively different time regimes have been also identified within a general
theoretical framework developed to analyze memory effects in granular compaction
\cite{ByP01}, although it has not been experimentally verified yet to the best
of our knowledge.

The energy of the one-dimensional Ising model in the configuration
$\bm{\sigma} \equiv \{ \sigma_{i} \}, \sigma_{i} = \pm 1$,  is
\begin{equation}
\label{2.1} {\cal H}(\bm{\sigma})=-J \sum_{k} \left( \sigma_{k}
\sigma_{k+1} -1 \right),
\end{equation}
where $J>0$ is the ferromagnetic coupling constant between nearest neighbour
spins. The Glauber dynamics for the model is formulated by means of a master
equation, with the transition rate from  configuration $\bm{\sigma}$ to
configuration $R_{k}\bm{\sigma}$, differing from $\bm{\sigma}$ only in the flip
of spin $\sigma_{k}$, given by \cite{Gl63}
\begin{equation}
\label{2.2} \omega_{k}(\bm{\sigma})=\frac{\alpha}{2} \left[
1-\frac{\gamma}{2} \sigma_{k} (\sigma_{k-1}+\sigma_{k+1}) \right].
\end{equation}
Here
\begin{equation}
\label{2.3} \gamma=\tanh \frac{2J}{k_{B}T},
\end{equation}
$k_{B}$ being Boltzmann's constant and $T$ the temperature of the system. The
quantity $\alpha$ determines the global time scale of the system. Although it
can be taken as temperature independent, in the context of structural glasses,
where spin configurations are associated with the minima of the potential energy
\cite{KyS90}, an Arrhenius expression
\begin{equation}
\label{2.4} \alpha(T)=\alpha_{0} \exp \left(- \frac{B}{k_{B}T}
\right),
\end{equation}
is often considered \cite{Sch88,ByP94}. The constant $B$ is a measure of the
energy barrier separating energy minima in configuration space and $\alpha_{0}$
can be employed to define a dimensionless basic time scale. In the following, we
will  use eq.\ (\ref{2.4}) for the sake of generality. The particular case of
temperature independent $\alpha$ can be obtained by taking the limit  $B
\rightarrow 0$.

We will restrict ourselves to homogeneous situations, i.e. states where all the
averages are translationally invariant. It is useful to introduce the spin-spin
correlations
\begin{equation}
\label{2.5} C_{n}(t) \equiv \langle \sigma_{k} \sigma_{k+n}
\rangle_{t} = \sum_{\bm{\sigma}} \sigma_{k} \sigma_{k+n} p \left(
\bm{\sigma},t \right),
\end{equation}
where $p(\bm{\sigma},t)$ is the probability of configuration
$\bm{\sigma}$ at time $t$. In an infinite lattice, the time
evolution of these correlations at constant temperature is given
by \cite{Gl63}
\begin{equation}
\label{2.6} \frac{d}{dt}C_{n}=-2 \alpha C_{n}+ \alpha \gamma
\left( C_{n-1}+C_{n+1} \right),
\end{equation}
for $ n \geq 1$, while $C_{0}=1$.  The equilibrium correlations are
\begin{equation}
\label{2.7} C_{n}^{(s)}(T)= \xi(T)^{n}, \quad \xi(T)=\tanh
\frac{J}{k_{B}T} \, .
\end{equation}
Denoting by $N$ the number of spins, a dimensionless energy per
spin can be defined as
\begin{equation}
\label{2.10} E(t)\equiv \frac{\langle {\cal H}
\rangle_{t}}{NJ}=1-C_{1}(t).
\end{equation}
An equation for the time evolution of $E(t)$ follows by
particularizing eq.\ (\ref{2.6}) for $n=1$. We will write it as an
expression for the instantaneous energy relaxation rate $r(t)$,
\begin{equation}
\label{2.11} r(t) \equiv -\frac{dE(t)}{dt}=\alpha (T) \left[
\mu_{1}(t)-\varepsilon(T) \mu_{2}(t) \right],
\end{equation}
where
\begin{equation}
\label{2.12} \mu_{1}(t)=2 E(t)-1+C_{2}(t) \geq 0, \quad
\mu_{2}(t)= 1+C_{2}(t) \geq 0,
\end{equation}
and we have introduced
\begin{equation}
\label{2.14} \varepsilon(T)=1-\gamma (T).
\end{equation}
Both $\varepsilon(T)$ and $\alpha(T)$ are monotonic increasing functions of the
temperature. An equation having the same structure as eq.\ (\ref{2.11}) was
proposed  to describe compaction in tapped granular systems  in ref.
\cite{ByP01}. The main difference is that here we have considered the time
evolution of the energy, instead of the density. In the interpretation of the
Ising model  as describing structural relaxation, an increase of the density
corresponds to a decrease of the energy. This is the reason why we have defined
$r(t)$ with a minus sign in eq.\ (\ref{2.11}).

Let us  suppose we carry out the following experiment. Starting from a given
configuration, the system is allowed to relax at constant temperature $T$. At a
certain time $t_{w}$, before the system has reached the steady configuration
corresponding to $T$, the temperature is suddenly changed to $T+\Delta T$. This
produces a  jump $\Delta r_{w}$ in the relaxation rate. For $\Delta T
\ll T$, keeping only first order terms in the perturbation $\Delta
T$, it is easily obtained that
\begin{equation}
\label{2.15} \Delta r_{w}= \lambda (t_{w}) \Delta \alpha,
\end{equation}
with
\begin{equation}
\label{2.16} \lambda(t)= \frac{r(t)}{\alpha}-\alpha
\mu_{2}(t)\frac{d\varepsilon}{d \alpha}=
\frac{r(t)}{\alpha}-\frac{\eta (1-\gamma^2)}{2} \mu_2(t) \, ,
\end{equation}
and $ \eta=4J/B$. The  function $\lambda(t)$ is defined along the relaxation
curve at constant T. As long as $\Delta T$ is small enough so that the linear
approximation we are considering is accurate, the function $\lambda(t)$ for
$t=t_{w}$ determines the relative behaviour of the relaxation rate jump $\Delta
r_{w}$ with respect to $\Delta
\alpha$ (or $\Delta T$).

To study the function $\lambda (t)$ we have to evaluate the relaxation rate
$r(t)$ and the correlation function $\mu_{2}(t)$ for a process at constant
temperature. The general solution of the hierarchy (\ref{2.6}) for an arbitrary
initial condition can be analytically found \cite{Gl63,BSyO70}. We will consider
processes starting in the state of maximum disorder, i.e. the equilibrium state
at infinite temperature. Similar results are obtained for other initial
conditions, as long as they correspond to highly disordered states. A simple
calculation gives \cite{Gl63,BSyO70}
\begin{equation}
\label{2.21} \Delta_{n}(t)\equiv C_n(t)-C_n^{(s)}(T)=
 - \frac{\gamma}{\pi} \int_{0}^{\pi}
dq\, \frac{\sin q \sin nq}{1-\gamma \cos q}\, e^{-2 \alpha t(1-
\gamma \cos q)}.
\end{equation}
Use of the above expression with $n=1$ into eq.\ (\ref{2.10})
yields
\begin{equation}
\label{2.22} E(t)-E^{(s)}(T)=\frac{\gamma}{\pi} \int_{0}^{\pi}
dq\, \frac{\sin^{2} q }{1-\gamma \cos q}\, e^{-2 \alpha t(1-
\gamma \cos q)},
\end{equation}
where $E^{(s)}(T)=1-C_1^{(s)}(T)$ is the equilibrium value of the dimensionless
energy per particle. The relaxation rate of the energy, defined in eq.\
(\ref{2.11}), for the present process is
\begin{equation}
\label{2.23} r(t) = \frac{2\alpha \gamma}{\pi} \int_{0}^{\pi} dq\,
\sin^{2} q \, e^{-2\alpha t(1- \gamma\cos q)} =\frac{e^{-2\alpha
t} I_{1}(2\alpha \gamma t)}{t} \, ,
\end{equation}
where  $I_{1}$ is the modified Bessel function of the first kind
\cite{AyS65}.

To evaluate $\mu_{2}(t)$ we use again eq.\ (\ref{2.21}), now for
$n=2$. In this way we get from eq.\ (\ref{2.12})
\begin{equation}
\label{2.24} \mu_{2}(t)=1+\xi^{2}-\frac{\gamma}{\pi} f(t),
\end{equation}
where
\begin{equation}
\label{2.25} f(t)=\int_{0}^{\pi} dq\, \frac{\sin q \sin
2q}{1-\gamma \cos q}\, e^{-2\alpha t(1-\gamma \cos q)}.
\end{equation}

Substitution of eqs.\ (\ref{2.23}) and (\ref{2.24}) into eq.\
(\ref{2.16}) leads to
\begin{equation}
\label{2.26} \lambda (t)= \frac{e^{-2 \alpha t} I_{1}(2 \alpha
\gamma t)}{\alpha t}- \frac{\eta (1-\gamma^{2} )}{2} \left[
1+\xi^{2} -\frac{\gamma}{\pi}\, f(t) \right].
\end{equation}
In the following we will consider that the temperature $T$ is very
low, so that $\varepsilon \ll 1$, $\gamma \rightarrow 1$, $\xi
\rightarrow 1$, and $\alpha \rightarrow 0$. It could be concluded
that the second term on the right hand side of eq.\ (\ref{2.26}) can be
neglected as compared with the first one \cite{note1} in this limit .
Nevertheless, this is not true in general,  and the relative relevance of both
terms depends on the time scale of interest. In particular, for $t \rightarrow
\infty$ the first term vanishes, as well as $f(t)$, and
\begin{equation}
\label{2.27} \lambda_{\infty} \equiv \lim_{t\rightarrow \infty}
\lambda (t)= -\frac{\eta (1-\gamma^{2})(1+\xi^{2})}{2} \sim -2
\eta \varepsilon < 0.
\end{equation}
Since $\lambda (t)$ is a continuous function of $t$, the above result implies
that, for large enough $t_{w}$, $\lambda (t_{w})<0$ and, therefore, $\Delta
r_{w} $ and  $\Delta\alpha$ have opposite signs. A sudden increase of the
temperature produces a negative jump of the relaxation rate. This is the kind of
``anomalous'' response found in compaction processes of granular materials
\cite{Ni99,JTMyJ00,ByP01}.

It is convenient to define the function
\begin{equation}
\label{2.28} \Lambda (t)=\lambda(t)-\lambda_{\infty}= \frac{e^{-2
\alpha t} I_{1}(2 \alpha \gamma t)}{\alpha t} +\frac{\eta \gamma
(1-\gamma^{2})}{2 \pi}\, f(t),
\end{equation}
that vanishes for $t \rightarrow \infty$. We are going to analyze the behaviour
of this quantity both in the short and the large time regions. First, consider
``short'' times characterized by
\begin{equation}
\label{2.28a} \alpha t \sim O(1),
\end{equation}
so that $\varepsilon \alpha t \ll 1$ for the low temperatures we are dealing
with. On the time scale defined above, we can approximate $\gamma$ by unity in
the expression of $f(t)$, eq.\ (\ref{2.25}), obtaining
\begin{equation}
\label{2.29} f(t)=\pi e^{- 2\alpha t} \left[ I_{0}(2 \alpha t)+2
I_{1} (2 \alpha t)+ I_{2} (2 \alpha t) \right].
\end{equation}
When this expression is substituted into eq.\ (\ref{2.28}), it is easily
realized that the term involving $f(t)$ can be neglected because of the $\eta
\varepsilon$ factor. Then,
\begin{equation}
\label{2.30} \Lambda(t) \sim \frac{e^{-2\alpha t} I_{1}(2 \alpha
t)}{\alpha t},
\end{equation}
for $\alpha t \sim O(1)$, $\varepsilon \ll 1$. In this time window,
$\lambda (t) \sim \Lambda (t) >0$, where we have taken into
account that $\lambda_{\infty}$ is of order $\varepsilon$ (see eq.\
(\ref{2.27})). As a consequence, the response of the system to a
small change $\Delta T$ in the temperature at $t=t_{w}$ is
``normal'' if $\alpha \varepsilon t_{w} \ll 1$. The modification of
the relaxation rate has the same sign as the temperature change.
Above, it was shown that for asymptotically large times the
response was qualitatively different. Let us study in more detail
what happens in the ``large'' time region. We define a slow time
scale by
\begin{equation}
\label{2.32} \tau \equiv \alpha \varepsilon t \sim O(1),
\end{equation}
i.e., times of the order of the system relaxation time. For these
times, it is $\alpha t = O(\varepsilon^{-1}) \gg 1 $ and we can
approximate $I_{1} (2\alpha \gamma t)$ by its asymptotic
behaviour for large argument \cite{AyS65}, getting
\begin{equation}
\label{2.33} \frac{e^{-2\alpha t} I_{1}(2 \alpha \gamma t)}{\alpha
t} \sim \varepsilon^{3/2} \frac{e^{-2 \tau}}{2 \pi^{1/2} \tau^{3/2}}.
\end{equation}
This quantity is much smaller than $\lambda_{\infty}$. Moreover, a
simple asymptotic analysis gives
\begin{equation}
\label{2.34} f(t) \sim (2\pi \varepsilon)^{1/2} \Gamma \left(
-1/2,2\tau \right),
\end{equation}
$\Gamma(a,x)$ being the incomplete Gamma function \cite{AyS65}.
Use of eqs. (\ref{2.33}) and (\ref{2.34}) into eq.\ (\ref{2.28})
leads to
\begin{equation}
\label{2.35} \Lambda(t) \sim \varepsilon^{3/2} \left[ \frac{e^{-2
\tau}}{2 \pi^{1/2} \tau^{3/2}} +\eta \left( \frac{2}{\pi}
\right)^{1/2} \Gamma \left( -1/2,2 \tau \right) \right],
\end{equation}
valid for $\varepsilon \ll 1$, $\tau =O(1)$. In this long time
regime, $ \Lambda (t)$ is of the order of $\varepsilon^{3/2}$, and
$|\lambda_{\infty}| \gg \Lambda(t)$. Then, $\lambda (t) \sim
\lambda_{\infty} <0$. The response of the system is anomalous, as
we already knew from the discussion of the asymptotic behaviour
for $t\rightarrow \infty$.

Since we have found that $\lambda (t)$ has different signs in the short and long
time regions, because of continuity reasons there must be at least one critical
time $t_{c}$ at which $\lambda (t_{c})=0$. This time separates regions with
normal and anomalous behaviour. Furthermore, the above discussion indicates that
this time must belong to an  intermediate time scale between those previously
considered, namely those given by eqs. (\ref{2.28a}) and (\ref{2.32}),
respectively. The matching time range where the two solutions, eqs.\
(\ref{2.30}) and (\ref{2.35}), are valid is characterized by
\begin{equation}
\label{2.36} \alpha t \gg 1, \quad \tau=\varepsilon \alpha t \ll 1.
\end{equation}
On this time scale, we can substitute the Bessel function in eq.\
(\ref{2.30}) by its asymptotic behaviour for large argument
\cite{AyS65}, with the result
\begin{equation}
\label{2.37} \Lambda (t) \sim \frac{1}{2 \pi^{1/2} (\alpha t
)^{3/2}}.
\end{equation}
The same expression is obtained from eq.\ (\ref{2.35}) in the
appropriate limit, showing that the solutions corresponding to the
short and large time scales match in the intermediate scale
defined by eq.\ (\ref{2.36}). A single uniform approximation
$\Lambda_{unif}(t)$, valid for all times, is obtained by adding
eqs. (\ref{2.30}) and (\ref{2.35}), and subtracting the expression
in the matching region, eq.\ (\ref{2.37}),
\begin{equation}
\label{2.38} \Lambda_{unif}(t)=  \frac{e^{-2\alpha t}
I_{1}(2\alpha t)}{\alpha t}+ \frac{e^{-2\alpha \varepsilon
t}-1}{2\pi^{1/2} (\alpha t)^{3/2}} +\varepsilon^{3/2} \eta \left(
\frac{2}{\pi} \right)^{1/2} \Gamma \left( -1/2,2\alpha \varepsilon t
\right).
\end{equation}

\begin{figure}
\twofigures[scale=0.4,clip=]{fig1.eps}{fig2.eps}
\caption{Comparison of the numerical evaluation of eq.\
(\protect{\ref{2.28}}) (circles) with the uniform asymptotic approximation given
by eq.\ (\protect{\ref{2.38}}) (solid line). Also plotted (dotted line) is the
asymptotic  long time value $|\lambda_\infty|$. Note that the quantities
represented in both axes are dimensionless.}
\label{fig1}
\caption{The dimensionless critical time $t_{c}$, separating the regions of
normal and anomalous response, as a function of the inverse of the reduced
temperature. The circles have been obtained by numerical evaluation of eq.\
(\protect{\ref{2.28}}), while the solid line is the result of the asymptotic
calculation, eq.\ (\protect{\ref{2.39}}).} \label{fig2}
\end{figure}

In fig.\ \ref{fig1} we compare this result with the exact expression, eq.\
(\ref{2.28}) for $\varepsilon =0.01$ and $J=B$, i.e. $\eta=4$. A very good
agreement is observed for all times. The relevant physical quantity we are
interested in is the critical time $t_{c}$, verifying $\lambda (t_{c})=0$ or,
equivalently, $\Lambda (t_{c})= -\lambda_{\infty}$. This latter quantity is also
indicated in the figure. Since $t_c$ is expected to belong to the intermediate
scale, we get from eqs. (\ref{2.27}) and (\ref{2.37})
\begin{equation}
\label{2.39} \alpha t_{c} \sim \left( 4 \pi^{1/2} \eta \varepsilon
\right)^{-2/3}.
\end{equation}
This result confirms that $t_{c}$ is in the overlapping region in agreement with
our previous ansatz, since in the limit $\varepsilon
\rightarrow 0$ it is $\alpha t_{c} \gg 1$ and $\tau_{c}= \varepsilon
\alpha t_{c} \ll 1$. Therefore, eq.\ (\ref{2.39}) is consistent
with the uniform expression (\ref{2.38}). It is interesting to
analyze the temperature dependence of $t_{c}$. By taking
logarithms in eq.\ (\ref{2.39}) we find for low $T$,
\begin{equation}
\label{2.40} \ln (\alpha_{0} t_{c})=-\frac{2}{3} \ln \left( 8
\pi^{1/2} \eta  \right) +\frac{B}{k_{B}T} \left( 1+\frac{2\eta}{3}
\right),
\end{equation}
showing an Arrhenius dependence. In fig.\ \ref{fig2} the logarithm of the
critical time $t_{c}$, given by eq.\ (\ref{2.39}), is plotted as a function of
the inverse of the reduced temperature $B/k_{B}T$, again for $\eta =4$. For
comparison, also the numerical values obtained from the exact expression, eq.\
(\ref{2.28}), are included. A linear profile, consistent with eq.\ (\ref{2.40}),
is observed.

Let us now address the particular case of $\alpha$ being temperature independent
that can be obtained by considering the limit $B\rightarrow 0$. Then $\eta$ goes
to infinity and $t_{c}$ vanishes. Therefore, the system always exhibits an
anomalous response to a temperature jump, independently of the time instant in
which it is produced. The physical reason for this result is clear, since eq.\
(\ref{2.15}) reduces in this situation to
\begin{equation}
\label{2.42} \frac{\Delta r_{w}}{\Delta T}=-\alpha \mu_{2w}
\frac{d\varepsilon}{dT}<0.
\end{equation}
Equation (\ref{2.42}) reflects that, if $\alpha$ does not depend on the
temperature, the rates of the processes decreasing the energy do not change
instantaneously in a temperature jump. This can be directly verified, for
instance, from eq.\ (\ref{2.11}).

In conclusion, the major result obtained in this paper is the analytical
derivation of memory effects in the relaxation of the one-dimensional Ising
model with Glauber dynamics. Moreover, these effects are qualitatively similar
to those observed experimentally in granular compaction
\cite{Ni99,JTMyJ00,ByP01}, for which the existence of a critical time separating
the regions of ``normal'' and ``anomalous'' responses, has been theoretically
argued \cite{ByP01}. Therefore, also in the present context the Ising model
seems to be the simplest relevant system beyond mean-field descriptions. The
results in this paper seems to indicate that the memory effects and the presence
of a critical time, as discussed here, may be quite general. In the Ising model,
the critical time lies in an intermediate time window, being much smaller than
the average relaxation time of the system towards its equilibrium state.

\acknowledgments
We acknowledge partial support from the
Direcci\'{o}n General de Investigaci\'{o}n Cient\'{\i}fica y
T\'{e}cnica (Spain) through Grant No. PB98-1124.

\end{document}